\documentclass[%
 twocolumn,
 amsmath,amssymb,
 aps,
]{revtex4-2}

\usepackage{color}

\usepackage{graphicx}% Include figure files
\usepackage{dcolumn}% Align table columns on decimal point
\usepackage{bm}% bold math
\begin{document}

\preprint{APS/123-QED}

\title{Feshbach resonances in an ultracold $^{7}$Li-$^{133}$Cs Bose-Bose mixture}% Force line breaks with \\
%\thanks{A footnote to the article title}%

\author{W.-X. Li}
\author{Y.-D. Chen}
\author{Y.-T. Sun}
\author{S.~Tung}
\email{stung@phys.nthu.edu.tw}
 % \email{Second.Author@institution.edu}
\affiliation{%
 Department of Physics, National Tsing Hua University, Hsinchu 30013, Taiwan\\
  and Center for Quantum Technology, Hsinchu 30013, Taiwan
}%

\date{\today}% It is always \today, today,
             %  but any date may be explicitly specified
\author{Paul S. Julienne}
\affiliation{Joint Quantum Institute, NIST and the University of Maryland, 100 Bureau Drive, Stop 8423, Gaithersburg, Maryland 20899-8423, USA}

\begin{abstract}
We present a study of interspecies Feshbach resonances in ultracold $^{7}$Li-$^{133}$Cs Bose-Bose mixtures. We locate ten interspecies resonances in three different spin-state combinations. By comparing to coupled-channel calculations, we assign six of the resonances to $s$-wave channels and the rest to $p$-wave channels. We use the $s$-wave resonances to refine the ground-state potentials of LiCs in the coupled-channel calculations and then obtain an accurate characterization of the scattering and bound-state properties of the mixtures. Our results will be useful for future experiments with ultracold $^{7}$Li-$^{133}$Cs mixtures.

\end{abstract}

\maketitle

%----------------------------------------------------------------------------------
\section{\label{sec:level1}Introduction}
Feshbach resonances can be exploited to control the interactions of ultracold atoms, providing a versatile platform to explore few-body and many-body phenomena \cite{RevModPhys.82.1225, RevModPhys.80.885}. Notable examples include molecular Bose-Einstein condensates \cite{10.1038/nature02199, 10.1126/science.1093280, PhysRevLett.91.250401}, Efimov states \cite{10.1038/nature04626, PhysRevLett.103.043201, 10.1016/j.aop.2006.10.011, 10.1088/1361-6633/aa50e8}, polarons \cite{PhysRevLett.102.230402, 10.1038/nature11065, PhysRevLett.117.055301, 10.1103/PhysRevLett.117.055301}, and unitary quantum gases \cite{10.1038/nature08814, 10.1126/science.1195219, 10.1126/science.1214987, 10.1038/nphys2850}. 

Feshbach resonances can also be used to produce Feshbach molecules~\cite{10.1103/RevModPhys.78.1311, 10.1103/PhysRevLett.97.180404, 10.1103/PhysRevLett.98.200403, PhysRevLett.97.120402, PhysRevLett.109.085301}. These loosely bound molecules can be transferred efficiently into ground-state molecules via stimulated Raman adiabatic passage. The scheme was first demonstrated in KRb molecules by Ni {\it et al.}~\cite{10.1126/science.1163861} and recently a dipolar quantum gas of KRb molecules was created by De Marco {\it et al.}~\cite{10.1126/science.aau7230}. The quantum gases of polar molecules provide unique opportunities to simulate many-body models with a strong anisotropic long-range interaction. Mixtures of Li and Cs have gained special interest, because fully polarized ground-state LiCs molecules possess a dipole moment almost ten times larger than KRb~\cite{doi:10.1063/1.1903944}. 

$^6$Li and $^7$Li are the stable isotopes of Li. The Feshbach resonances of $^6$Li and $^{133}$Cs have been studied in Refs.~\cite{PhysRevA.87.010701, PhysRevA.87.010702} and further analyzed in Ref.~\cite{PhysRevA.90.012710}. Recently, a few $^7$Li-$^{133}$Cs Feshbach resonances have been predicted based on a coupled-channel (CC) model using {$^6$Li-$^{133}$Cs} singlet and triplet potentials~\cite{naidon2020magnetic}. In this paper we report the observation of ten {$^7$Li-$^{133}$Cs} Feshbach resonances. These resonances are identified by enhanced trap losses of Li as the scanning magnetic field hits a resonance. Furthermore, we carry out full CC calculations to provide a characterization of the Li-Cs interactions. 

We organize this article as follows. In Sec.~II we describe the experimental setup and procedure for the trap-loss measurements. We also introduce a model to analyze finite-temperature atom-loss features. In Sec.~III we show the results from our coupled-channel calculations. In Sec.~IV we conclude with a discussion of the features of the Feshbach resonances and give an outlook on the future applications.

%----------------------------------------------------------------------------------
\section{Experiment and Analysis}

The experimental setup is based on the apparatus detailed in Ref. \cite{PhysRevA.103.023102}, which employs a lithium-cesium slow beam. The setup now includes two overlapping magneto-optical traps (MOTs) and two separated optical dipole traps, one MOT and one dipole trap for each species. The $^{7}$Li MOT captures approximately $10^8$ Li atoms from the slow beam in 40~s, and the temperature of the Li atoms in the MOT is $\sim$0.4~mK. For $^{133}$Cs, we use a dark-spot MOT \cite{PhysRevLett.70.2253} to suppress losses caused by light-assisted collisions \cite{10.1007/s100530050576}. The Cs MOT collects $10^{6}$ Cs atoms in a loading time of 2~s. Following a compression phase in the MOTs, the Cs atoms are further cooled down to $5$~$\mu$K using gray molasses cooling \cite{PhysRevA.98.033419}. %Before loading each species into its optical dipole trap, we optically pump the Cs atoms to the $\vert F=3, m_{F}=3\rangle$ state and the Li atoms to the states of $F = 1$ manifold.   
\begin{figure}[h]
\centering
   \includegraphics[width=0.45\textwidth]{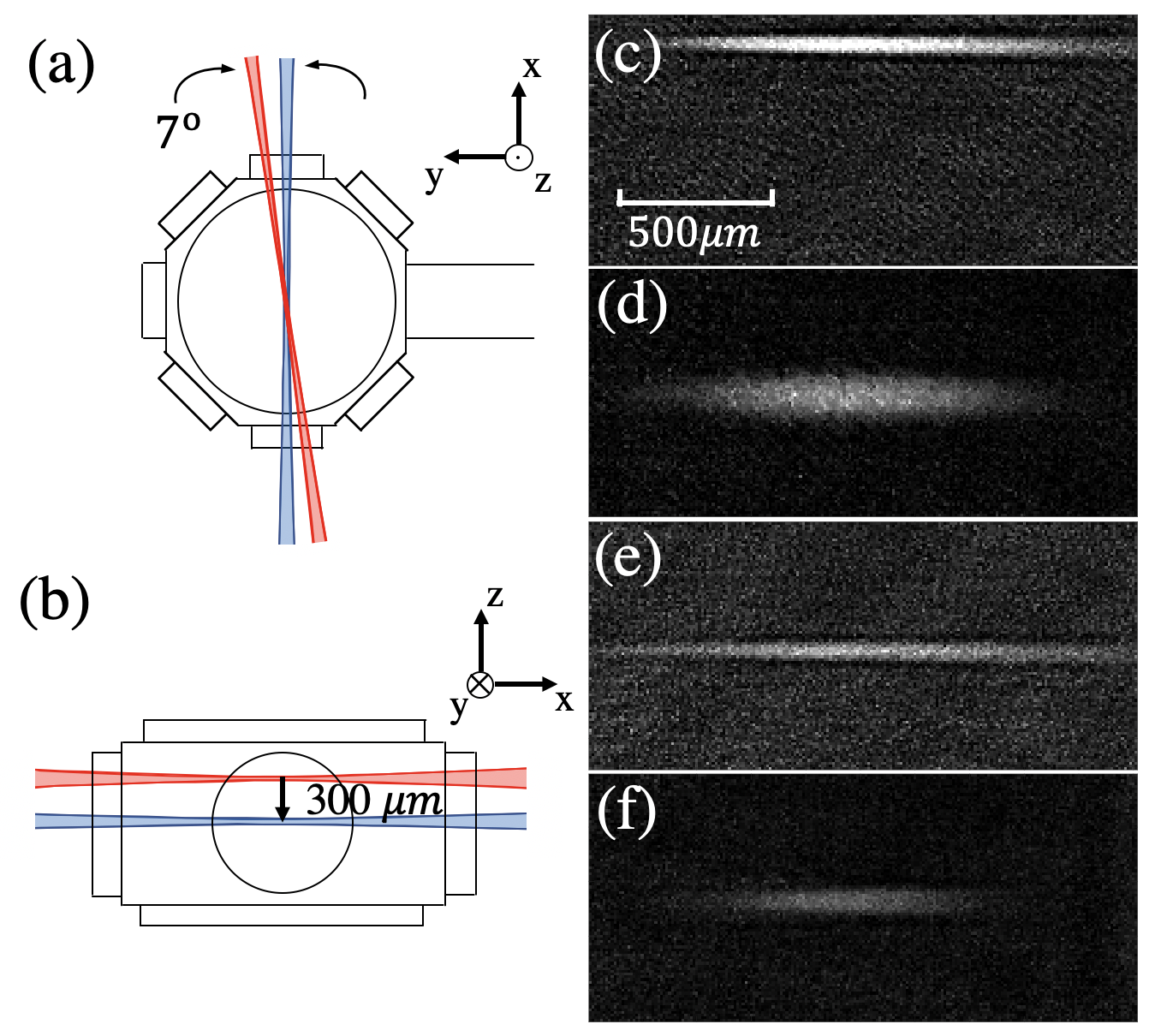}
   \caption{Schematic of the Li and Cs optical dipole traps (not to scale): (a) top view and (b) side view.  The single-beam optical dipole trap of Li (red) and Cs (blue) cross at the center of the cell with an angle of 7$^{\circ}$. The two dipole traps are initially displaced to achieve the maximum loading. After the MOT loading, we move the Li trap down to merge with the Cs trap. Also shown are absorption images of (c) Li and (d) Cs taken in separate traps and (e) Li and (f) Cs taken after the merge.}
   \label{fig:Setup}
\end{figure}

%The two MOTs have a vertical separation of 300~$\mu$m, optimized for loading Li and Cs atoms 
To prepare ultracold $^{7}$Li-$^{133}$Cs mixtures for trap-loss measurements, we load the Li and Cs atoms into separated dipole traps (one for each species) first, then perform evaporation, and finally merge the two traps. The dipole traps are single-beam traps, created by independent 1064-nm laser sources. At this wavelength, the polarizability of Cs is approximately four times greater than that of Li. The two dipole beams cross at the center of the science cell with an angle of 7$^{\circ}$, as shown in Fig.~\ref{fig:Setup}. They are initially separated by a vertical distance of 300~$\mu$m. The distance is chosen such that the loading efficiencies of the two species are optimized. We use a high-power tightly focused ($P$ = 160~W and $\omega_0$ = 50~$\mu$m) beam for Li and a low-power loosely focused ($P$ = 6~W and $\omega_0$ = 130~$\mu$m) beam for Cs. The trap depths are estimated to be 2~mK for Li and 50 $\mu$K for Cs.

Independent evaporative cooling procedures are implemented for Li and Cs. Before evaporation, we apply a bias magnetic field of 819~G and a field gradient of 31.1~G/cm. The bias field enables efficient Li evaporation and the field gradient allows the Cs atoms in the $\vert F = 3, m_F = 3\rangle$ state to fully levitate against gravity. The Cs evaporation is carried out by reducing the gradient linearly to 0 in 2.7~s. In the absence of the field gradient, the gravitational pull reduces the Cs trap depth by $\sim$50$\%$. The Li evaporation is implemented by two exponential power ramp downs. Each ramp takes  1.2~s and the two ramps are separated by a hold time of 0.3~s. The Li trap depth reduces to 12$\%$ of the initial value after the first ramp and further to 6$\%$ after the second ramp.  

The two traps are merged to allow interactions between Li and Cs. During the merge, the Li beam controlled by an acousto-optic deflector is moved down to overlap with the Cs beam. At the same time, the optical power of the Li beam is reduced by a factor of 30 to mitigate the heating from loading the large, cold Cs clouds into the tight Li trap. The Cs trap is adiabatically turned off once the two traps overlap, and the optical power of the Li beam is increased slightly to enhance the Li-Cs interactions. The process takes 150~ms, and in the end, we trap $2\times10^5$ Li atoms and $10^5$ Cs atoms with temperatures of 8 and 60~$\mu$K, respectively, in the same dipole trap. 

\begin{figure}[h]
\centering
   \includegraphics[width=0.5\textwidth]{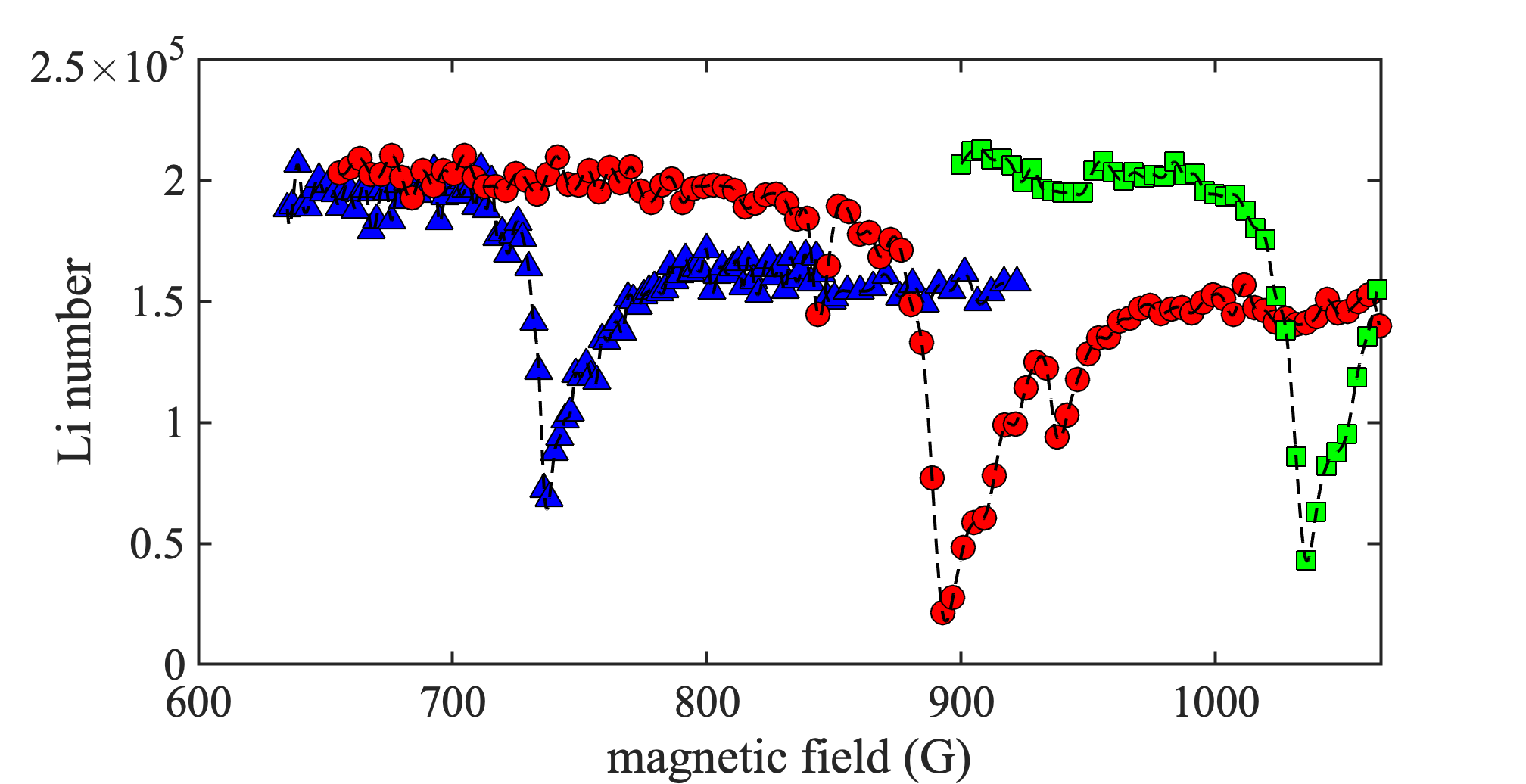}
   \caption{Feshbach spectroscopy for the Li$\vert a \rangle$ (blue triangles), Li$\vert b \rangle$ (red circles), and Li$\vert c \rangle$ (green squares) states. We find a total number of five resonances: one at 738~G for Li$\vert a \rangle$, three at 844, 893, and 938~G for Li$\vert b \rangle$, and one at 1036~G for Li$\vert c \rangle$. Dashed lines are a guide to the eye.}
   \label{fig:LiFBR}
\end{figure}

We explore the Li-Cs Feshbach resonances in three different scattering channels: Li$\vert a \rangle$ + Cs$\vert a \rangle$, Li$\vert b \rangle$ + Cs$\vert a \rangle$, and Li$\vert c \rangle$ + Cs$\vert a \rangle$, where Li$\vert a \rangle$ = Li$\vert F = 1, m_F = 1 \rangle$, Li$\vert b \rangle$ = Li$\vert F = 1, m_F = 0 \rangle$, Li$\vert c \rangle$ = Li$\vert F = 1, m_F = -1 \rangle$, and Cs$\vert a \rangle$ = Cs$\vert F = 3, m_F = 3 \rangle$. At high fields, a basis set of the electronic and the nuclear angular momentum projections $\vert m_J,m_I\rangle$ becomes more suitable for Li; thus, Li$\vert a \rangle$ = {$\vert m_J = -\frac{1}{2}, m_I = \frac{3}{2} \rangle$}, Li$\vert b \rangle$ = {$\vert m_J = -\frac{1}{2}, m_I = \frac{1}{2} \rangle$}, and Li$\vert c \rangle$ = {$\vert m_J = -\frac{1}{2}, m_I = -\frac{1}{2} \rangle$}. Here Li$\vert a \rangle$ and Cs$\vert a \rangle$ are the lowest Zeeman levels.

To facilitate the assignment of Feshbach resonances, we prepare both species in the selected spin states for Feshbach (trap-loss) spectroscopy. We use optical pumping to prepare Cs in the Cs$\vert a \rangle$ state. We confirm that $85\%$ of the Cs atoms are pumped to the designated state by the absorption images after Stern-Gerlach separations; nevertheless, only Cs$\vert a \rangle$ atoms remain in the trap after the Cs evaporation. To prepare the Li spins, we pump the Li atoms to the $F$ = 1 manifold and populate all three spin states. After the Li evaporation, most of the remaining atoms are in the Li$\vert b \rangle$ state. The Li atoms in the Li$\vert a \rangle$ and the Li$\vert c \rangle$ states are depleted because of the inelastic spin-changing collisions \cite{PhysRevA.77.023604, J.Phys.B50.01LT01}. The collisions flip the spins of the colliding Li$\vert a \rangle$ and Li$\vert c \rangle$ atoms and put them in the Li$\vert b \rangle$ state. At $B=819$~G, following each of the collisions, an excessive energy of $k_\textrm{B} \times 1.4$~mK is deposited to the atoms. The acquired kinetic energies allow them to escape from the trap. To check the Li spin purity, we perform Feshbach spectroscopy on the Li-only samples. In the scans, we find three resonant loss features (see red circles in Fig.~\ref{fig:LiFBR}); two are known resonances in the Li$\vert b \rangle$-Li$\vert b \rangle$ channel~\cite{10.1016/j.crhy.2010.10.004} and the one at 938~G is close to the Li$\vert b \rangle$-Li$\vert c \rangle$ resonance~\cite{doi:10.1063/1.5131023}. From the profile, we estimate the spin purity of Li$\vert b \rangle$ to be~$\sim$90$\%$. 

The spin preparation of Li$\vert a \rangle$ and Li$\vert c \rangle$ is done through adiabatic rapid passage (ARP) from the Li$\vert b \rangle$ state. During the ARP, we keep the bias magnetic field at~600~G. At this field, the transition frequencies of Li$\vert b \rangle$ $\rightarrow$ Li$\vert a \rangle$ and Li$\vert b \rangle$ $\rightarrow$ Li$\vert c \rangle$ differ by 37~MHz; thus, it is possible to achieve a selective transfer. We also take the Li-only Feshbach spectroscopy after the selective transfers (see Fig.~\ref{fig:LiFBR}). For the samples transferred to Li$\vert a \rangle$, one resonant feature is found at 738~G. The position agrees well with the previous observations for the Li$\vert a \rangle$-Li$\vert a \rangle$ Feshbach resonance~\cite{PhysRevA.77.023604, 10.1016/j.crhy.2010.10.004}. We also found one resonant feature at 1036~G for the samples transferred to Li$\vert c \rangle$. The position is also in good agreement with the theoretical prediction for the Li$\vert c \rangle$-Li$\vert c \rangle$ Feshbach resonance~\cite{doi:10.1063/1.5131023}. 

To perform interspecies Feshbach spectroscopy, we ramp the bias magnetic field to a desired value and measure the atom number of $^7$Li after a variable hold time, which is adjusted empirically to give the best signal-to-noise ratio. The remaining Li atoms are measured by standard absorption imaging at zero magnetic field. Near an interspecies Feshbach resonance, the Li samples experience enhanced trap losses through the three-body recombination processes, cross thermalization, and inelastic two-body collisions. A total of ten resonance features are located within the magnetic field range of $300-1065$ G. These features do not exist when the Cs atoms are absent. They are fitted with Gaussian functions, from which the resonance positions~$B^{\textrm{expt}}_{\textrm{0}}$ and widths~$\delta^{\textrm{expt}}$ are determined. The results are summarized in Table~~\ref{tab:summary}. We calibrate the magnetic field using the microwave transition between Cs$\vert F = 3, m_F = 3 \rangle$ and Cs$\vert F = 4, m_F = 3 \rangle$.

The resonant loss features, including the loss peak positions and widths, are temperature dependent. At a temperature of a few $\mu$K, thermal effects are usually negligible; the loss peaks shift from their zero-energy resonance positions by no more than 0.1~G. Since our measurements are taken at higher temperatures, $T_{\textrm{Cs}}$ = $60-100$~$\mu$K, several thermal effects would need to be taken into account in order to achieve a precise determination of resonance position. First, all $s$-wave resonances observed in this work are closed-channel dominated resonances with $s_{\textrm{res}}<1$ \cite{RevModPhys.82.1225}. In this case, the shift of pole position is approximately $E/ \delta \mu$. Here $E$ represents the collision energy and is equal to $k_B T$; $\delta \mu$ is the difference between the magnetic moment of the separated atoms and the magnetic moment of the bare bound state. Since $\delta \mu > 0 $ for all the observed resonances, the resonance positions shift to higher fields at higher temperatures. Second, thermal averaging over a broad collision energy distribution will also shift the peaks. Finally, the intrinsic width of resonance increases as $E^{1/2}$ for $s$ waves and $E^{3/2}$ for $p$ waves \cite{10.1088/0953-4075/33/5/201}. Thermally broadened line shapes of different $m_l$ components from a $p$-wave resonance can overlap strongly with one another, leading to an excessive error in the determination of the resonance position. Therefore, to avoid the errors, we perform the optimization of the Li-Cs potentials using only the six $s$-wave resonances. We note that the assignment of the resonances is done by comparing the measured resonance positions to the pole positions predicted by the CC calculations (see Sec.~III).

\begin{figure}[b]
\centering
   \includegraphics[width=0.5\textwidth]{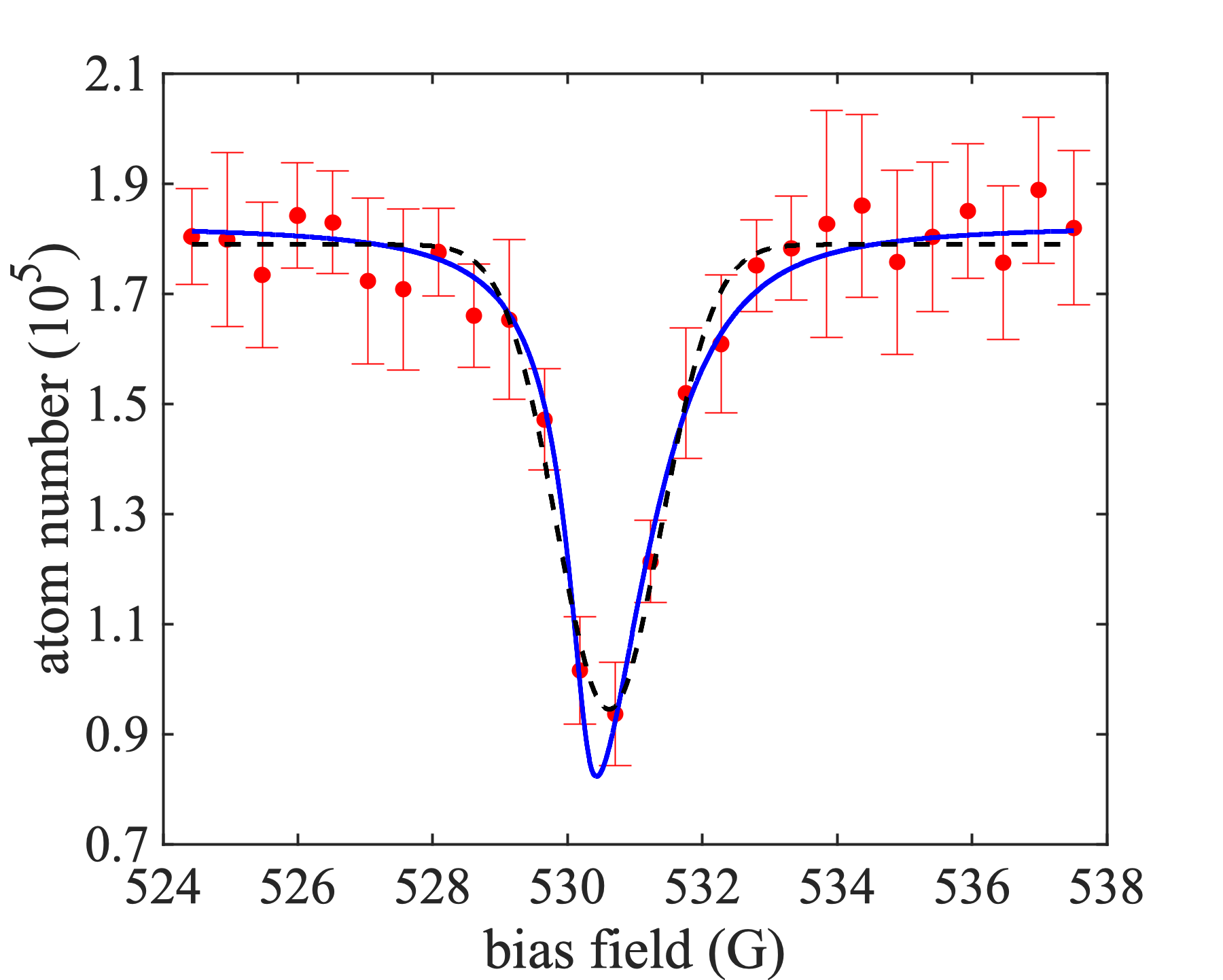}
   \caption{Li trap loss data from the $aa$ channel near the $s$-wave resonance.  The dashed line is a Gaussian fit to the data, and the loss leak position is located at 530.6~G. The solid line is another fit done with the procedure described in the Appendix. In the fitting procedure, the temperature $T$, resonance position $B_0$, decay rate $\gamma$, and initial Li number $N_{\textrm{Li}}(0)$ are the free parameters. The resonance position $B_0=530.3$~G is found in the fit.  }
   \label{fig:profile_fit}
\end{figure}

Exploiting the analogy to optical Feshbach resonance and photoassociation~\cite{PhysRevA.92.022709}, a similar theoretical model is developed to describe an $s$-wave Feshbach resonance with decay to loss channels (Appendix). The loss channels include, but are not limited to, cross thermalization and decay of quasimolecules~\cite{PhysRevLett.120.193402, 10.1103/PhysRevA.99.052704}. Based on the model, we can derive a thermally averaged loss rate coefficient with chosen parameters and then integrate the rate equation for the Li number to generate an atom-loss line shape for fitting. The zero-energy resonance positions can be determined from the finite-temperature model fits. Figure~\ref{fig:profile_fit} shows the fit for the $s$-wave resonance of the Li$\vert a \rangle$ + Cs$\vert a \rangle$ channel.
The fitted resonance positions of all $s$-wave resonances are given in Table~\ref{tab:summary}. 

\begin{table*}[t]
\caption{Summary of the $^7$Li-$^{133}$Cs Feshbach resonances. The resonance positions $B^{\textrm{expt}}_{0}$ and widths $\delta^{\textrm{expt}}$ are determined by Gaussian fits to the atom-loss features. The errors in $B^{\textrm{expt}}_{0}$ and $\delta^{\textrm{expt}}$ come from the fits, representing a 95\% confidence level. We also note that the field calibration and drift result in an  error of 0.5~G. The $B_0^{\textrm{fit}}$ are the zero-energy resonance positions determined by the finite-temperature model fits. The theoretical values for resonance positions $B^{\textrm{CC}}_{0}$ and zero crossing $B^{\textrm{ZC}}$ are obtained from the coupled-channel calculations with the refined $^7$Li-$^{133}$Cs potentials. The errors in $B^{\textrm{CC}}_{0}$ include the uncertainties from the fits, the field calibration and drift, and the thermal effects.  }
\centering
\begin{tabular*}{\textwidth}{@{\extracolsep{\stretch{1}}}*{8}{c}@{}}
\hline\hline
\\[-1 em]
&\multicolumn{6}{c}{This work}&$\textrm{Ref.~\cite{naidon2020magnetic}}$\\
\cline{2-7}
\\[-1em]
Channel & $B^{\textrm{expt}}_{0} (\textrm{G})$ &$\delta^{\textrm{expt}} (\textrm{G})$&$l$& $B^{\textrm{fit}}_0$&$B^{\textrm{CC}}_{0} (\textrm{G})$&$B^{\textrm{ZC}}(\textrm{G})$ & $B_{0} (\textrm{G})$\\
\hline  
\\[-1em]
$^{7}$Li$|1, 1\rangle$ + $^{133}$Cs$|3, 3\rangle$ 	&353.2(4) &0.7(2)&1&& &  &  \\ 
$^{7}$Li$|1, 1\rangle$ + $^{133}$Cs$|3, 3\rangle$    &439.8(2) &1.9(3)&1&& &  &   \\
$^{7}$Li$|1, 1\rangle$ + $^{133}$Cs$|3, 3\rangle$    &530.6(1) &1.1(1)&0&530.3&530.2(6)& 534.5 &538.43 \\
\\[-0.5 em]
$^{7}$Li$|1, 0\rangle$ + $^{133}$Cs$|3, 3\rangle$ 	&561.8(1) &2.0(2)&1&& & & \\
$^{7}$Li$|1, 0\rangle$ + $^{133}$Cs$|3, 3\rangle$  &610.0(2) &2.6(3)&0&609.8&609.7(6) & 620.1 &  618.31  \\
$^{7}$Li$|1, 0\rangle$ + $^{133}$Cs$|3, 3\rangle$  &732.3(3) &3.0(5)&0&732.0&731.9(7) & 738.6  &  735.76 \\

\\[-0.5 em]												
$^{7}$Li$|1, -1\rangle$ + $^{133}$Cs$|3, 3\rangle$ 	&631.8(2) &1.8(3)&1&& & &  \\
$^{7}$Li$|1, -1\rangle$ + $^{133}$Cs$|3, 3\rangle$    &717.4(2) &3.5(4)&0&717.1&717.1(6) &  733.7&   \\
$^{7}$Li$|1, -1\rangle$ + $^{133}$Cs$|3, 3\rangle$   &779.7(2) &1.4(2)&0&779.2&779.1(7) & 780.9 &   \\
$^{7}$Li$|1, -1\rangle$ + $^{133}$Cs$|3, 3\rangle$   &850.8(7) &5.0(10)&0&850.6&850.7(9) & 858.1 &  \\
\hline\hline
\end{tabular*}
\label{tab:summary}
\end{table*}

%----------------------------------------------------------------------------------

\section{Theory}
We take advantage of two open-source programs, {\footnotesize BOUND} \cite{BOUND} and {\footnotesize MOLSCAT} \cite{MOLSCAT}, to implement CC calculations for this work. The {\footnotesize BOUND} program solves the Schr\"{o}dinger equation for the bound states and obtains the binding energies at a fixed value of magnetic field \cite{doi.org/10.1016/j.cpc.2019.02.017}. The {\footnotesize MOLSCAT} program carries out the scattering calculations for the scattering lengths \cite{doi.org/10.1016/j.cpc.2019.02.014}. Either {\footnotesize BOUND} or {\footnotesize MOLSCAT} can be used to locate Feshbach resonances. The calculations are done with a fully decoupled basis set $\vert s_{\textrm{Li}}, m_{s,\textrm{Li}}\rangle\vert i_{\textrm{Li}}, m_{i,\textrm{Li}}\rangle\vert s_{\textrm{Cs}}, m_{s,\textrm{Cs}}\rangle\vert i_{\textrm{Cs}}, m_{i,\textrm{Cs}}\rangle\vert L,M_L\rangle $. Here $s$ and $i$ represent the electron and nuclear spins, respectively, and $L$ is the relative angular momentum of the two atoms. 

To use the programs for the $^{7}$Li-$^{133}$Cs mixtures, we construct the CC model based on the $^{6}$Li-$^{133}$Cs work \cite{PhysRevA.87.010702}. First, we perform a mass-scaled version of the calculation and obtain resonance positions that are systematically higher ($\sim5$~G) than the observed values. We then add a small harmonic variation $\alpha_i(R-R_e)^2$ to the inner wall of each potential in the region $R<R_e$, where $R_e$ is the equilibrium internuclear distance, $\alpha$ is the adjustable parameter, and $i$ is the potential index. We assign $i = 1$ and $3$ for the singlet and triplet potentials, respectively. The two parameters $\alpha_1$ and $\alpha_3$ are adjusted to make slight changes in the singlet and triplet scattering lengths $a_1$ and $a_3$, and the resulting field values of the resonances are used to fit the observed $B$ fields of the resonances. 

We use the fitted resonance positions $B_0^{\textrm{fit}}$ from the finite-temperature model fits to optimize the $^{7}$Li-$^{133}$Cs potentials in the CC calculations. The optimization is implemented using {\footnotesize BOUND}, and the theoretical values of the resonance positions are given in Table~\ref{tab:summary}. All molecular states associated with the $s$-wave resonances have mixed singlet-triplet characters and zero orbital angular momentum projection along the internuclear axis; thus, the projection of the total angular momentum $m_F$ is the only good quantum number. The {\footnotesize MOLSCAT} program is used to calculate the scattering lengths for the three channels at different magnetic fields. The refined potentials of $^{7}$LiCs yield singlet scattering length $a_1 = 45.82(2)a_0$ and triplet scattering length $a_3=873.8(70)a_0$. These values can be compared with $a_1 = 45.47a_0$ and $a_3=908.2a_0$, reported in Ref.~\cite{naidon2020magnetic}. The large $a_3$ indicates the presence of a near-threshold triplet bound state.

\begin{figure}[t]
\centering
   \includegraphics[width=0.5\textwidth]{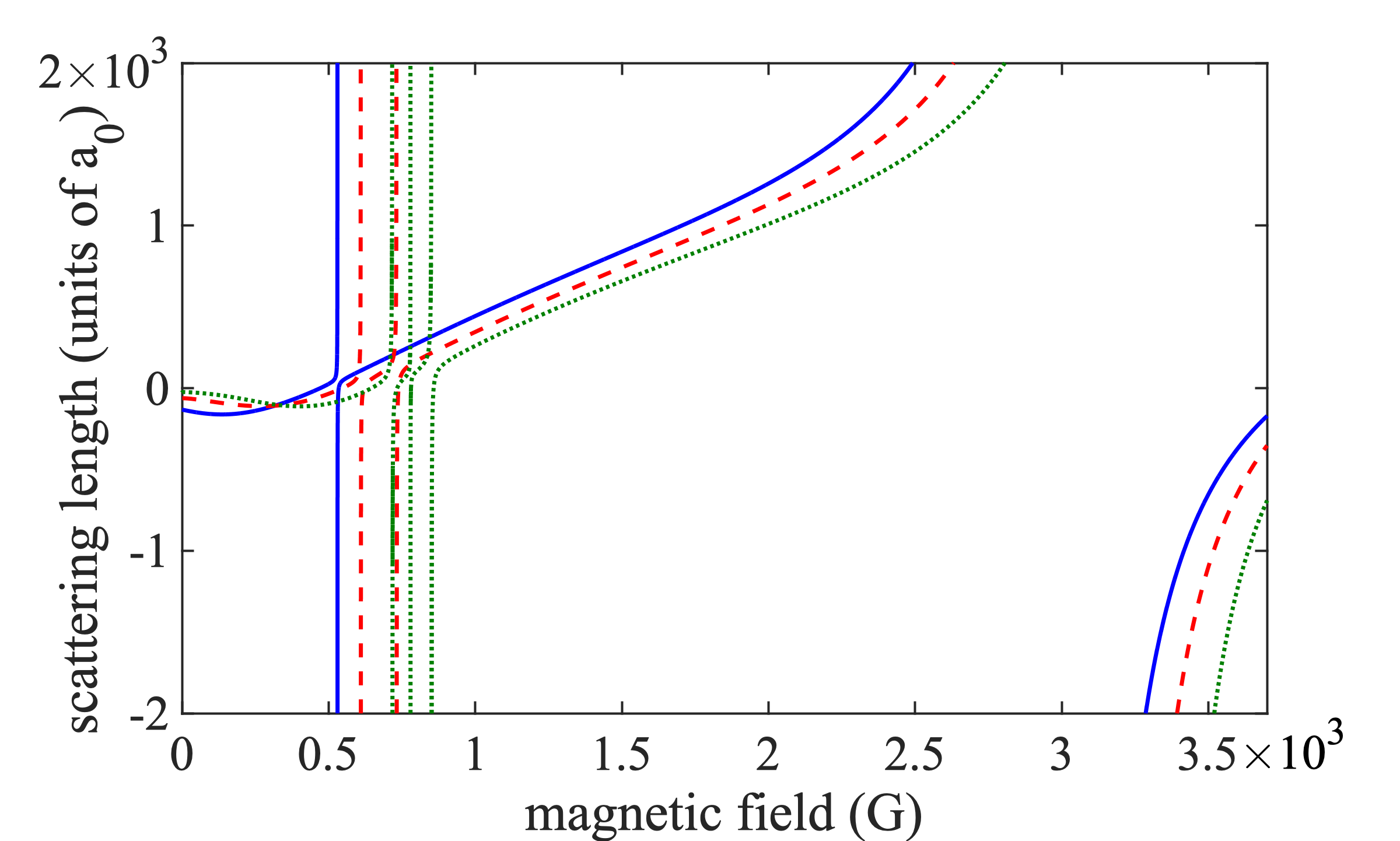}
   \caption{Theoretical values for scattering lengths of the Li$\vert a \rangle$ + Cs$\vert a \rangle$ (blue solid line),  Li$\vert b \rangle$ + Cs$\vert a \rangle$(red dashed line), and Li$\vert c \rangle$ + Cs$\vert a \rangle$ (green dotted line) channels. The resonances located at $B<1000~\textrm{G}$ are strongly influenced by strong Feshbach resonances near 3000~G, causing scenarios of overlapping Feshbach resonances.}
   \label{fig:a_largeB}
\end{figure}

\begin{figure*}
\centering
   \includegraphics[width=1\textwidth]{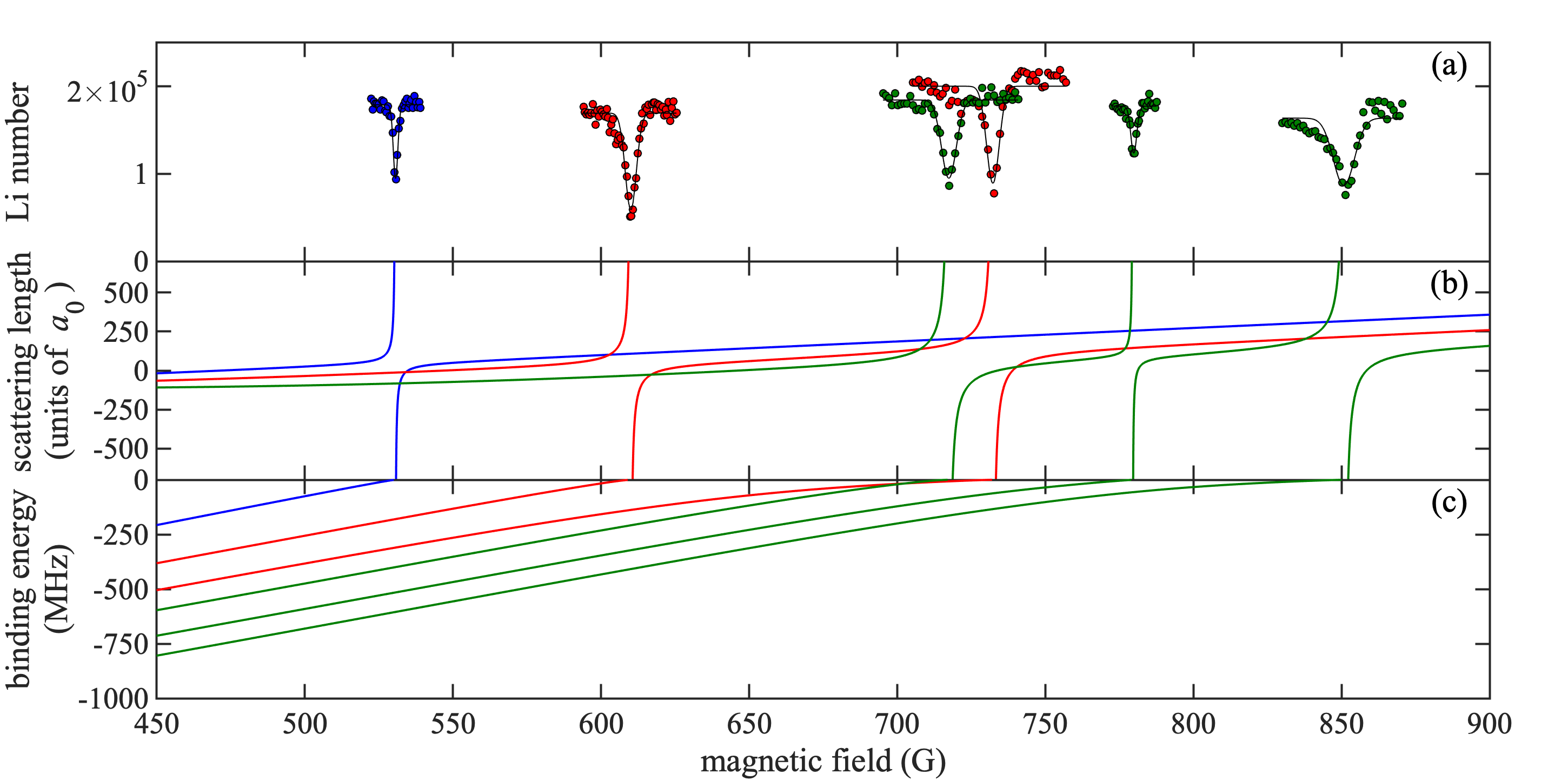}
   \caption{Feshbach spectroscopy and the coupled-channel calculations for the $s$-wave resonances. (a) Observed Li trap loss features in the Li$\vert a \rangle$ + Cs$\vert a \rangle$ (blue),  Li$\vert b \rangle$ + Cs$\vert a \rangle$ (red), and  Li$\vert c \rangle$ + Cs$\vert a \rangle$ (green) channels. The solid lines are the Gaussian fits to the atom-loss features. Also shown are the theoretical values of (b) scattering lengths and (c) binding energies. }
   \label{fig:LiCsFBR}
\end{figure*}

The overlapping effect are really pronounced in the $s$-wave resonances. In each channel, there is a nearly linear increase of the background scattering length with $B$, associated with the presence of extremely strong resonances located around 3000~G (see Fig.~\ref{fig:a_largeB}). The single pole approximation,  i.e.,~$a(B) =a_\mathrm{bg} \left[1- \Delta / (B-B_{0}) \right]$, only becomes valid  within a small field range near the resonances. To summarize the CC calculations, we plot the measured atom-loss features for the $s$-wave resonances with the scattering lengths and binding energies in Fig.~\ref{fig:LiCsFBR}. The calculated pole and zero crossing positions are given in Table~\ref{tab:summary}.

	%----------------------------------------------------------------------------------
\section{Conclusion and discussion}

We have conducted a detailed study on the $^{7}$Li-$^{133}$Cs Feshbach resonances. We used Feshbach spectroscopy to locate ten interspecies Feshbach resonances in three different scattering channels. We also carried out the coupled-channel calculations to provide precise characterization of the interaction parameters, such as scattering length and binding energy. From the results, we are able to identify two interesting field regimes for the mixtures. First, near $B = 885$~G, the Li$\vert b \rangle$ + Cs$\vert a \rangle$ mixture has moderate inter- and intraspecies interactions; the respective values of the scattering lengths are approximately $250a_0$, $100a_0$, and $400a_0$ for the Li$\vert b \rangle$ + Cs$\vert a \rangle$, Cs$\vert a \rangle$ + Cs$\vert a \rangle$, and Li$\vert b \rangle$ + Li$\vert b \rangle$ channels. These values can support fast thermalization and slow three-body recombinations. Thus, it may be possible to create a dual-species Bose-condensed system of Li$\vert b \rangle$ and Cs$\vert a \rangle$ near 885~G. Second, near the Li$\vert b \rangle$ + Cs$\vert a \rangle$ resonance at $B = 732$~G, the Cs$\vert a \rangle$+Cs$\vert a \rangle$ scattering length is positive and large, $\sim$5000$a_0$, while the Li$\vert b \rangle$+Li$\vert b \rangle$ scattering length is positive and small, $\sim$9$a_0$. In this regime, Bose-Einstein condensates of Li$\vert b \rangle$ can interact resonantly with strongly interacting Cs atoms. 

Finally, radio-frequency spectroscopy of LiCs dimers in the future can provide accurate measurements on the dimer binding energies near a resonance. These measurements can set tighter constraints to the pole positions and hence lead to highly accurate scattering length data. In addition, the spectra can also be used to verify higher partial-wave ($l \geq1$) models in coupled-channel calculations.

%----------------------------------------------------------------------------------

\begin{acknowledgements}
The authors acknowledge Jos\'{e} P. D'Incao for fruitful discussions. This work was supported by the Ministry of Education, Taiwan and the National Science and Technology Council, Taiwan.
\end{acknowledgements}
%\nocite{*}

\appendix*
\section{}
In the presence of the decay that leads to atom loss, the $S$-matrix element describing an $s$-wave scattering channel coupled to an isolated bound state near a threshold is
\begin{equation}
S(k) =  \left\{ 1- \frac{i\hbar \Gamma(k)}{E- E_{\textrm{res}}+i \frac{1}{2}\hbar [\gamma+\Gamma(k)]}\right\} e^{2i\eta_{\textrm{bg}}},
\label{eqn:S_k}
\end{equation}
where $E_{\textrm{res}}=\delta\mu(B-B_0)$ is the energy of the resonance, $B_0$ is the resonance position at the scattering threshold ($E=0$), and $\gamma$ is the rate of atom loss. The decay rate $\gamma$ includes all loss sources. In the case of magnetic Feshbach resonances, $\hbar \Gamma(k) = 2ka_{\textrm{bg}}\delta\mu \Delta$ is related to the width of the resonance. The phase shift $\eta_{\textrm{bg}}$ is associated with the background scattering length $a_{\textrm{bg}}$ as $\tan\eta_{\textrm{bg}} = -k a_{\textrm{bg}}$.  

Taking the expression for $S(k)$, the inelastic cross section $\sigma_{\textrm{in}}=\pi[1-\vert S(k)\vert^2]/k^2$ can be written as
\begin{equation}
\sigma_{\textrm{in}} = \frac{2\pi}{k}\frac{\gamma^2 l_{\textrm{eff}}}{(\frac{E-E_{\textrm{res}}}{\hbar})^2+\frac{\gamma^2}{4}\left( 1+2k l_{\textrm{eff}}\right)^2},
\label{eqn:sigma_in}
\end{equation}
where $l_{\textrm{eff}}=a_{\textrm{bg}}\delta\mu\Delta/\hbar \gamma$ is a characteristic length associated with the coupling strength. Then the thermally averaged loss rate coefficient $\langle K\rangle_{\textrm{th}}$ is given by 
\begin{equation}
 \langle K\rangle_{\textrm{th}} = \sqrt{\frac{8}{\mu \pi}}\frac{1}{(k_b T)^{3/2}} \int dE \sigma(E) E e^{-E/k_b T}.
\label{eqn:K}
\end{equation}

The final step to derive an atom-loss line shape is to assume a loss model. In our experiment, the temperature of Cs is much higher than Li and the trap depth is much shallower ($\sim\frac{1}{4}$) for Li. Thus, the dominant loss of Li is heating from colliding with Cs. In this case, the loss rate equation for the Li number $N_{\textrm{Li}}$ is 
 
\begin{equation}
\frac{dN_{\textrm{Li}}}{dt} = - \langle K\rangle_{\textrm{th}} \overline{n}_{\textrm{Cs}} N_{\textrm{Li}},
\label{eqn:rate_equation}
\end{equation}
where $\overline{n}_{\textrm{Cs}} = \int n_{\textrm{Cs}}(\vec{r})n_{\textrm{Li}}(\vec{r})d^3\vec{r}/ \int n_{\textrm{Li}}(\vec{r})d^3\vec{r}$ is the average Cs density. In our measurements, the number loss of Cs is less than $10\%$; thus, we assume the Cs density is constant in the rate equation. After integrating Eq.~(\ref{eqn:rate_equation}) over an interaction time, we obtain atom-loss line shapes that can fit the measurements. Here $T$, $B_0$, $\gamma$, and $N_{\textrm{Li}}(0)$ are the free parameters in the fits.

\end{document}